\documentclass[twocolumn]{aastex631}

\hypersetup{linkcolor=red,citecolor=blue,filecolor=cyan,urlcolor=blue}

\usepackage{graphicx,color}
\usepackage{amssymb}
\usepackage{amsmath}
\usepackage{url}
\usepackage{natbib}
\usepackage{txfonts}
\usepackage{multirow}
\usepackage{array}
\usepackage{rotating}
\usepackage{sidecap}
\usepackage{hyperref}
\usepackage{epstopdf}
\usepackage{footnote}
\usepackage{tabularx}
\usepackage{booktabs}
\usepackage{longtable}
\usepackage{tabu}
\usepackage{longtable}

\newcommand{\MgII}{\ion{Mg}{2}}

\newcommand{\FeX}{\ion{Fe}{10}}
\newcommand{\FeIX}{\ion{Fe}{9}}

\newcommand{\kms}{km~s$^{-1}$}
\newcommand{\degree}{\ensuremath{^\circ}}

\newcommand{\sdo}{\textit{SDO}}
\newcommand{\iris}{\textit{IRIS}}
\newcommand{\hri}{HRI$_{EUV}$}

\shortauthors{Panesar et al}

\begin{document}
\title{ Solar Orbiter and SDO Observations, and Bifrost MHD Simulations of Small-scale Coronal Jets }


\correspondingauthor{Navdeep K. Panesar}
\email{panesar@lmsal.com}

\author[0000-0001-7620-362X]{Navdeep K. Panesar}
\affil{Lockheed Martin Solar and Astrophysics Laboratory, 3251 Hanover Street, Bldg. 252, Palo Alto, CA 94304, USA}
\affil{Bay Area Environmental Research Institute, NASA Research Park, Moffett Field, CA 94035, USA}

\author[0000-0003-0975-6659]{Viggo H. Hansteen}
\affil{Lockheed Martin Solar and Astrophysics Laboratory, 3251 Hanover Street, Bldg. 252, Palo Alto, CA 94304, USA}
\affil{Bay Area Environmental Research Institute, NASA Research Park, Moffett Field, CA 94035, USA}
\affil{Rosseland Centre for Solar Physics, University of Oslo, P.O. Box 1029 Blindern, NO–0315 Oslo, Norway}
\affil{Institute of Theoretical Astrophysics, University of Oslo, P.O. Box 1029 Blindern, NO–0315 Oslo, Norway}

	\author[0000-0001-7817-2978]{Sanjiv K. Tiwari}
\affil{Lockheed Martin Solar and Astrophysics Laboratory, 3251 Hanover Street, Bldg. 252, Palo Alto, CA 94304, USA}
\affil{Bay Area Environmental Research Institute, NASA Research Park, Moffett Field, CA 94035, USA}

\author[0000-0003-2110-9753]{Mark C. M. Cheung}
 \affil{Lockheed Martin Solar and Astrophysics Laboratory, 3251 Hanover Street, Bldg. 252, Palo Alto, CA 94304, USA}

	\author[0000-0003-4052-9462]{David Berghmans}
 \affil{Solar-Terrestrial Centre of Excellence – SIDC, Royal Observatory of Belgium, Ringlaan -3- Av. Circulaire, 1180 Brussels, Belgium}

\author[0000-0001-9027-9954]{Daniel M{\"u}ller}
\affil{European Space Agency, ESTEC, P.O. Box 299, 2200 AG Noordwijk, The Netherlands}

\begin{abstract}
	We report high-resolution, high-cadence observations of five small-scale coronal jets in an on-disk quiet Sun region observed with  Solar Orbiter's EUI/\hri\ in 174 \AA. We combine the \hri\ images with  the EUV images of SDO/AIA and investigate magnetic setting of the jets using co-aligned line-of-sight magnetograms from SDO/HMI. The \hri\ jets are miniature versions of typical coronal jets as they show narrow collimated spires with a base brightening. Three out of five jets result from a detectable minifilament eruption following flux cancelation at the neutral line under the minifilament, analogous to coronal jets. To better understand the physics of jets, we also analyze five small-scale jets from a high-resolution Bifrost MHD simulation in synthetic \FeIX/\FeX\ emissions.	The jets in the simulation reside above neutral lines and four out of five jets are triggered by magnetic flux cancelation. The temperature maps show the evidence of cool gas in the same four jets. Our simulation also shows the signatures of opposite Doppler shifts (of the order of $\pm$10s of \kms) in the jet spire, which is evidence of untwisting motion of the magnetic field in the jet spire. The average jet duration, spire length, base width, and speed in our observations (and in synthetic \FeIX/\FeX\ images) are 6.5$\pm$4.0 min (9.0$\pm$4.0 min), 6050$\pm$2900 km (6500$\pm$6500 km), 2200$\pm$850 km, (3900$\pm$2100 km), and 60$\pm$8 \kms\ (42$\pm$20 \kms), respectively.
Our observation and simulation results provide a unified picture of small-scale solar coronal jets driven by magnetic reconnection accompanying flux cancelation. This picture also aligns well with the most recent reports of the formation and eruption mechanisms of larger coronal jets. 


\end{abstract}



\section{Introduction} \label{sec:intro}

The Extreme Ultraviolet Imager (EUI; \citealt{rochus2020}), on board Solar Orbiter \citep{muller2020}, observed numerous small-scale brightenings, known as `campfires' \citep{berghmans2021}. Using 5-minutes of EUI data by the telescope EUV High Resolution Imager, \hri,  \cite{berghmans2021} reported that the duration of campfires ranges from 10 s to 200 s. They appear at coronal temperatures between 1 MK and 1.6 MK and at the bright edges of chromospheric network lanes. Using triangulation techniques, it was determined that some of the campfires are located between 1000 and 5000 km above the photosphere \citep{zhukov21}. However, the photospheric magnetic field environment and their evolution in relation with campfires was not known in their study. 

\cite{panesar2021} studied the dynamics and magnetic field evolution of 52 randomly selected campfires by combining EUI 174 \AA\ images together with EUV images from \sdo/AIA and line-of-sight magnetograms from \sdo/HMI. They found that (i) campfires are rooted at the edges of photospheric magnetic network lanes; (ii) most of the campfires reside above neutral lines and 77\% of them appear at sites of magnetic flux cancelation between the majority-polarity magnetic flux patch and a merging minority-polarity flux patch, with a flux cancelation rate of $\sim$10$^{18}$ Mx hr $^{-1}$; (iii) some of the smallest campfires come from the sites where magnetic flux elements were barely discernible in HMI; (iv) 79\% (41 out of 52) of campfires are accompanied by structures of cool plasma, analogous to small-scale filaments in coronal jets; and (iv) campfires resemble small-scale jets, loops, and dots, and some of them have a `complex' structure. They concluded that `campfire' is a general term that encompasses various types of small-scale solar features/brightenings. 

Later \cite{kahil22} also examined the line-of-sight photospheric magnetic field evolution of campfires using Solar Orbiter's PHI data and found similar results as of \cite{panesar2021}, in that majority of campfires are triggered by magnetic flux cancelation.

Some of the campfires show similarities with coronal jets \cite[e.g.][]{panesar16b,panesar18a,panesar2021,hou21}. They occur at the edges of quiet Sun network regions and are driven by the eruption of cool-plasma structure. Further, there are several other campfire-like events that were observed by \hri\ e.g. fine-scale dots in an emerging region \citep{tiwari22}, plasma jets \citep{chitta21}, and plasma flows \citep{mandal21,li22-EUI}.

Similar to EUI campfires, several small-scale tiny brightenings \cite[e.g.][]{alpert16}, low-lying loop nanoflares \citep{winebarger13}, small-scale loops/surges \citep{tiwari19}, and fine-scale jets (also known as jetlets; \citealt{panesar19}) have been observed by high resolution EUV images from Hi-C \citep{kobayashi14,rachmeler19}. Jetlets occur at the edges of magnetic network lanes  and they show similarities with coronal jets \citep{raouafi14,panesar18b,panesar20b}. 
Although erupting minifilament-like dark structures have not been seen at the base of jetlets.  

Using  3D radiation MHD simulation, \cite{chen21} reported that campfires are mostly caused by component reconnection between interacting bundles of magnetic field lines. In one out of seven events, they found the evidence of twisted flux rope in agreement with the observational findings of \cite{panesar2021}. But in their study \cite{chen21} did not provide any evidence of flux emergence or flux cancelation at the base of any of their seven analyzed campfires. 

Here, we present examples of five on-disk small-scale solar coronal jets observed by the telescope  \hri\ of EUI onboard Solar Orbiter together with EUV images from Solar Dynamics Observatory
(SDO)/Atmospheric Imaging Assembly (AIA; \citealt{lem12}), and line-of-sight magnetograms from SDO/Helioseismic and Magnetic Imager (HMI; \citealt{scherrer12}). We find that these small-scale jets are miniature version of larger coronal jets. The only difference is that these are four to five times smaller in size than typical coronal jets.
To better understand the physics  of coronal jets, we also synthesize \FeIX/\FeX\ emissions from a Bifrost magnetohydrodynamic (MHD) simulation. By investigating the plasma flows and  magnetic field evolution in the synthetic images, we find that coronal jets reside above magnetic neutral lines, where magnetic flux convergence and cancelation normally occurs, in agreement with the observations. They often show opposite Doppler shifts (blueshifts and redshifts) at  the jet spire during the eruption onset. Four out of five jets show the evidence of cool-plasma as we observe in the EUI and AIA images. 


\floattable
\begin{center}
	\begin{table*}[ht]
		\caption{Properties of Observed and Simulated Coronal Jets \label{tab:list}}
		\renewcommand{\arraystretch}{1.1}
		\begin{tabular}{c*{7}{c}}
			\noalign{\smallskip}\tableline\tableline \noalign{\smallskip}
			
			Event  &  Time  & Jet duration  & Jet spire length & Jet-base width  & Visibility of  & Flux & Speeds \\
			no.  & (UT)  & (min) & (km)   & (km) & Cool plasma & Cancelation &  (\kms) \\
				\noalign{\smallskip}\hline \noalign{\smallskip}
			Jets in the EUI Observations\scriptsize$^\star$  \\
			\noalign{\smallskip}\hline \noalign{\smallskip}
			1 &  21:20:52  & 2 $\pm$ 10s  & 3800 $\pm$ 325  & 1900 $\pm$ 200  & No  & A\tablenotemark{\scriptsize a}  &  47 $\pm$ 5.0 \\ 
			2 &  21:48:12  & 7 $\pm$ 10s  & 7650 $\pm$ 260  & 3700 $\pm$ 2450  & A\tablenotemark{\scriptsize b}   & A\tablenotemark{\scriptsize a} & 55 $\pm$ 9.0\\ 
			3&  22:13:02\tablenotemark{\scriptsize c}  & 13 $\pm$ 10s  & 10400 $\pm$ 810  & 1900 $\pm$ 780  & Yes  & Yes & 65 $\pm$ 10.0\\ 
			4&  22:05:02\tablenotemark{\scriptsize c}  & 8 $\pm$ 10s  & 5000 $\pm$ 160  & 2000 $\pm$ 225  & Yes  & Yes & 58 $\pm$ 0.5 \\ 
			5  &  21:38:22\tablenotemark{\scriptsize c}  & 3 $\pm$ 10s  & 3300 $\pm$ 350  & 1500 $\pm$ 240  & Yes  & Yes & 69 $\pm$ 5.0\\ 
			\noalign{\smallskip}\hline \noalign{\smallskip} 
			average$\pm$1$\sigma$$_{ave}$  & & 6.5 $\pm$ 4.0  & 6050 $\pm$ 2900 &  2200 $\pm$ 850 & &   & 59 $\pm$ 8.5\\
			\noalign{\smallskip}\tableline\tableline \noalign{\smallskip}
			Jets in the Simulation \\
			\noalign{\smallskip}\hline \noalign{\smallskip}
			1 &  t0 + 5000s & 13 $\pm$ 50s  & 11200 $\pm$ 390  & 5780 $\pm$ 110  & Yes & Yes &  43 $\pm$ 1.4\\ 
			2 &  t0 + 4900s & 6.6 $\pm$ 50s  & 5390 $\pm$ 445  & 3100 $\pm$ 160  & Yes & No flux &  40 $\pm$ 5.0\\ 
			3 &  t0 + 6850 & 9 $\pm$ 50s  & 4130 $\pm$ 120  & 1770 $\pm$ 80  & No  & Yes &  12 $\pm$ 2.5 \\   
			4 &  t0 + 3600 & 14 $\pm$ 100s  & 7500 $\pm$ 1900  & 6600 $\pm$ 400  & Yes & Yes & 72 $\pm$ 9.0  \\    
			5 &  t0 +4500  & 4 $\pm$ 50s  & 4400 $\pm$ 190  & 2160 $\pm$ 150  & Yes  & Yes &  40 $\pm$ 7.0\\   
			\noalign{\smallskip}\hline \noalign{\smallskip} 
			average$\pm$1$\sigma$$_{ave}$  & & 9.0 $\pm$ 4.0  & 6500 $\pm$ 2900 & 3900 $\pm$ 2100 & &  &  41.5 $\pm$ 21.0  \\
			\noalign{\smallskip}\tableline\tableline \noalign{\smallskip}
		\end{tabular}
		\singlespace
		\tablecomments{			
			\\\textsuperscript{a}{Ambiguous, cancelation between weak flux elements.}
			\\\textsuperscript{b}{Ambiguous, cool structure appears but does not erupt.}
			\\\textsuperscript{c}{These jets are taken from \cite{panesar2021}.
		\\\textsuperscript{$\star$}{Except the plane-of-sky speeds, which are estimated using AIA 171 \AA\ images.}}}
	\end{table*}		
\end{center}


\begin{figure*}[ht]
	\centering
	\includegraphics[width=\linewidth]{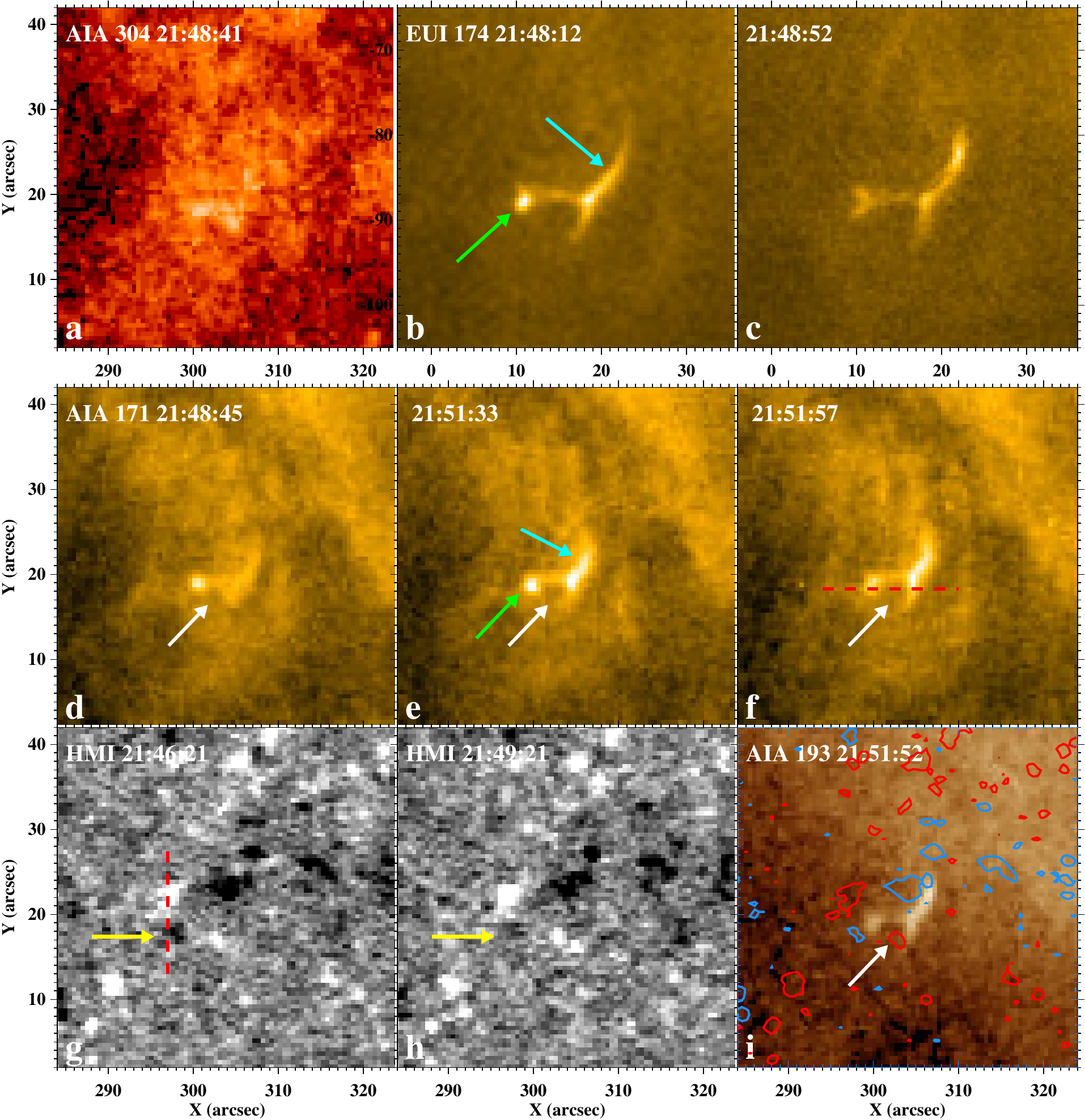} 
	\caption{Example-1 of small-scale coronal jet (event-2 of Table \ref{tab:list}). Panels (b) and (c) show the 174  \AA\ \hri\ images of the jet. Panels (a), (d--f), and (i) show the same jet in AIA 304, 171, and 193 \AA\ images, respectively.  The cyan and green arrows, respectively, point to the jet spire and jet base brightening. The white arrow points to the minifilament structure. Panels (g) and (h) display the line of sight photospheric magnetic field of the jet-base region. The yellow arrows point to jet-base minority-polarity negative magnetic flux patch that converge towards the neutral line. In panel (i), HMI contours, of levels $\pm$15 G, at 21:46:21 UT are overlaid, where red and blue contours outline positive and negative magnetic flux, respectively.  The red-dashed line in (f) and (g), respectively, show the cut for the  time–distance map of Figures \ref{fig2a}a and \ref{fig2a}b. The animation (Movie1a) runs from 21:40 to 21:54 UT, while the AIA animation (Movie1b) runs from 21:40 to 22:00 UT. Both the animations are unannotated and  FOV are same as in this Figure. 
		}  \label{fig1}
\end{figure*} 
%
%
%
\begin{figure*}
	\centering
	\includegraphics[width=\linewidth]{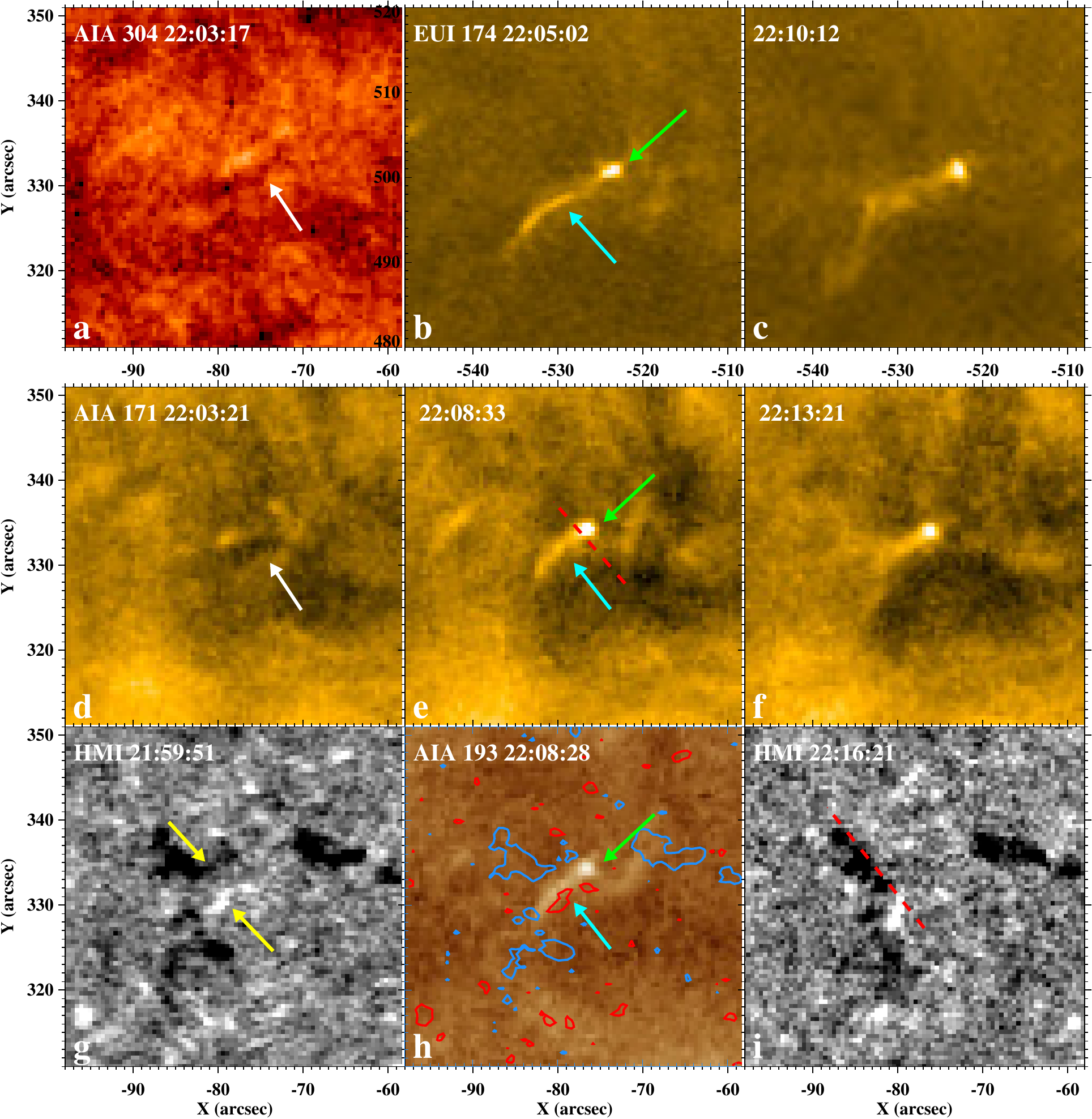}
	\caption{Example-2 of small-scale coronal jet (event-4 of Table \ref{tab:list}). Panels (b) and (c) show the 174  \AA\ \hri\ images of the jet. Panels (a), (d--f), and (h) show the same jet in AIA 304, 171, and 193 \AA\ images, respectively.  The cyan and green arrows, respectively, point to the jet spire and jet base brightening. The white arrow points to the minifilament structure. Panels (g) and (i) display the line of sight photospheric magnetic field of the jet-base region. The yellow arrows point to converging positive and negative flux patches where the minifilament resides. In panel (h), HMI contours, of levels $\pm$15 G, at 21:59:51 UT are overlaid, where red and blue contours outline positive and negative magnetic flux, respectively.  The red-dashed line in (e) and (i), respectively, show the cut for the  time–distance map of Figures \ref{fig2a}c and \ref{fig2a}d.  The animation (Movie2a) runs from 21:46 to 22:17 UT and the animation is unannotated, while the AIA animation (Movie2b) runs from 21:53 to 22:21 UT and the annotations and FOV are same as in this Figure. 
	}  \label{fig2}
\end{figure*} 
%
%
%
%
\section{DATA Sets}\label{data} 
\subsection{Solar Orbiter/EUI}

We randomly selected five small-scale coronal jets in a quiet region using data of  \hri\ on 20-May-2020\footnote{https://doi.org/10.24414/z2hf-b008}. The observations were taken when the instrument was in the commissioning phase.  The \hri\ wavelength passband is centered on 174 \AA\ and detects the emission from \FeIX\ and \FeX\ lines that form at around 1 MK. The HRI of EUI captured high resolution images (pixel size of 0\arcsec.492; \citealt{rochus2020}) at high temporal cadence (10 s). The Solar Orbiter was positioned at 0.612 au from the Sun on 20-May-2020, thus the \hri\ images have a pixel size of 217 km. During this period of observation (between 21:20 and 22:17 UT) \hri\ captured 5 images at a 10 s cadence plus a 6th image 70 s later. These observations were made as a part of a technical compression test of \hri, therefore having a variable settings. Nonetheless, 60 images are well exposed, un-binned, and compressed at high quality levels \citep{panesar2021,tiwari22}. 
The \hri\ events appear 3.22 minutes earlier than the SDO/AIA images because, aforementioned, Solar Orbiter was closer to the Sun than the SDO. The separation angle between Solar Orbiter and SDO was 16.33\degree.

\subsection{SDO/AIA and SDO/HMI}

We also use extreme ultraviolet (EUV) images captured with AIA onboard SDO. AIA provides full-Sun images with high spatial
resolution (1.5\arcsec) and high temporal cadence
(12 s) in seven EUV wavelength bands \citep{lem12}. For our present study, we mainly use images from 304, 171, and 193 \AA\ channels because the jets are best seen in these three wavelengths. To investigate the photospheric magnetic field of the jet-base regions, we use line-of-sight magnetograms from HMI \citep{scherrer12}. In addition to other data products, HMI provides full-disk line of sight (LOS) magnetograms at 45 s temporal cadence with pixels of 0.5\arcsec. The HMI magnetograms have a noise level of about $\pm$7 G \citep{schou12,couvidat12}. We summed two consecutive magnetograms to enhance the visibility of weak magnetic flux patches. 

The AIA and HMI datasets have been downloaded from JSOC\footnote{http://jsoc.stanford.edu/ajax/exportdata.html} cutout service and co-aligned using SolarSoft routines \citep{freeland98}.  JHelioviewer software was also used to locate the jets in SDO/AIA \citep{muller17}. 

\subsection{Bifrost Simulation}
We analyze a Bifrost magnetohydrodynamic (MHD) simulation of a quiet Sun network region  to better interpret  the structure and dynamics of coronal jets and the observations. For more details of this particular simulation, see  papers by \cite{pontieu22} and \cite{tiwari22}.
A horizontal flux sheet of varying strength is injected at the bottom boundary of a model that spans a domain of 72$\times$72$\times$61 Mm$^{3}$ on a grid of [720, 720, 1115] using the Bifrost code \citep{gudiksen2011}. The model reaches 8.5 Mm below the photosphere and then extends up to 52.5 Mm into the corona,  above the photosphere.  The scattering in the chromosphere with optically thick radiative transfer is included in the Bifrost model \citep[e.g.][]{skartlien2000,hayek2010}. The model includes optically thin radiative transfer in the middle chromosphere to the corona following the methods described by \cite{carlsson2012}. The equation of state (together with partial ionization of the atmospheric plasma) is treated in local thermodynamic equilibrium (LTE) through a look-up table that is constructed using solar abundances.

Importantly, thermal conduction along the magnetic field, particularly relevant for the solar corona, is included in the energy equation. We synthesize Fe IX and Fe X emissions by integrating the contribution function  ${\phi(u,T) n_{\rm e} n_{\rm H} G(T,n_{\rm e}})$ along the line of sight.
Here, $\phi(u,T) = 1/\sqrt(\pi) \Delta\nu_D  \exp[-((\Delta\nu-\nu  u.n/c)/ \Delta \nu_D)^2]$, where $\Delta\nu = \nu-\nu_0$, with $\nu_0$ being the laboratory frequency of the transition involved, u.n is the velocity along the line of sight. The thermal broadening profile corresponds to a width $\Delta\nu_D = \nu_0/c \sqrt(2kT/m_A)$, where $m_A$ is the mass of the radiating atom, and T is the temperature. The values are obtained along the line of sight from the simulation and then the contribution function is integrated along the line of sight for every frequency point \citep{hansteen10}. The ionization and excitation states of the emitting ions are taken from CHIANTI  \citep{dere97,young03}.
For more details see papers by \cite{hansteen10,hansteen15}. Since the wavelength of the iron lines is short, there is the possibility of absorption from neutral gas.  We include this effect in our calculation by multiplying the contribution function along the line of sight with $\exp(-\tau)$ where $\tau$ is the combined opacity of  hydrogen and helium, as well as from singly ionized helium (see \citealt{pontieu09} for details). 

We analyzed five small-scale jets from the simulations. We mainly use synthetic \FeIX\ and \FeX\ lines for our jet study because  \FeIX\ and \FeX\ lines are the main lines of \hri\ 174 \AA\ passband. We note that the EUI 174 \AA\ passband also includes emission from oxygen lines formed at roughly 280,000--320,000 K. Such emission will affect our synthetic spectra, but trial calculations indicate that they will not significantly change our analysis. Further, we also averaged synthetic \FeIX\ and \FeX\ lines and create a single map \FeIX/\FeX, in order to maintain the similarity with the \hri\ observations.  Similarly, we also average two Doppler maps (for \FeIX\ and \FeX) to create a single Doppler map. To study the magnetic field evolution of jets in simulation we use Bz maps (vertical component of the magnetic field) as the same field of view of \FeIX/\FeX\ maps and Doppler maps. 
These are large scale simulation with a pixel size of 100 km and a temporal cadence of 50 s.

\section{Results}
\subsection{Overview}\label{over}
We examine the dynamics and evolution of five on-disk small-scale coronal jets observed by Solar Orbiter/EUI and SDO/AIA. Their jet-base photospheric magnetic field structure is examined by using line of sight magnetograms from SDO/HMI. Further, to better understand the dynamics and magnetic field evolution of small-scale jets and to better interpret the observations, we studied five coronal jets from a high resolution Bifrost magnetohydrodynamic (MHD) simulation. In Section \ref{cjet}, we show two examples of EUV jets from the observations and in Section \ref{bjet}, we present two example jets from the Bifrost MHD simulation. Table \ref{tab:list} shows the measured parameters of all our jets observed by Solar Orbiter/EUI and found in the simulation.

\subsection{Solar Orbiter and SDO observations of coronal jets}\label{cjet}

 \subsubsection{Example-1}
 
In Figure \ref{fig1}, we display a jet example-1 (event-2 of Table \ref{tab:list}) observed by EUI and AIA. The corresponding animations (Movie1a and Movie1b) show the same images over the duration of the event. The jet is clearly visible in AIA 171 and 193 \AA\ images but it does not show-up as clearly as in AIA 304 \AA\ image (Figure \ref{fig1}a). 

\begin{figure*}
	\centering
	\includegraphics[width=\textwidth]{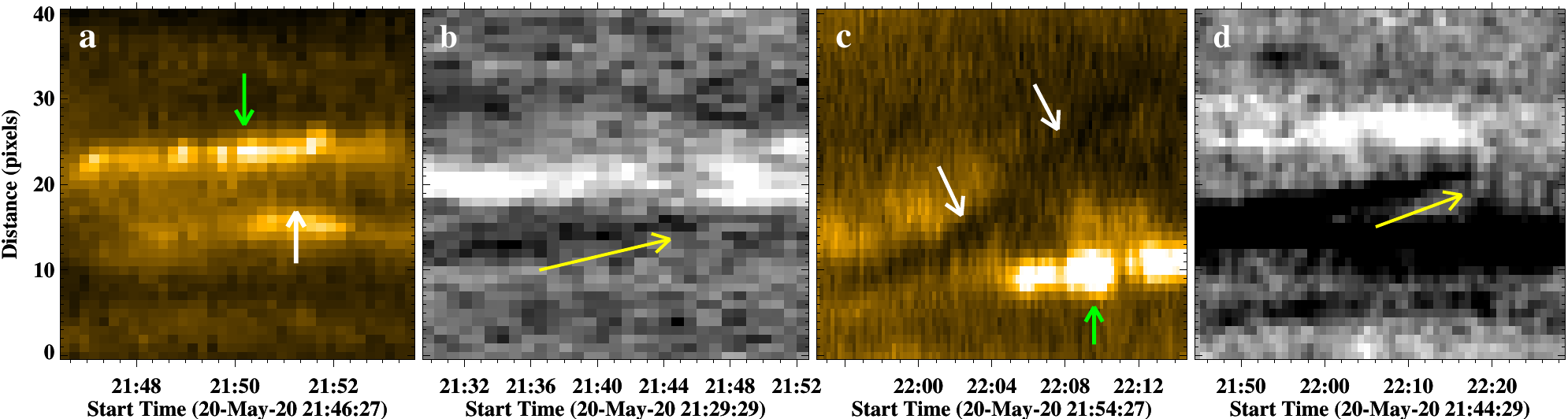}
	\caption{Panels (a) and (b), respectively, show the AIA 171 \AA\ intensity and HMI time-distance maps of the minifilament and magnetic field  of the jet of Figure \ref{fig1}. AIA 171 \AA\ and HMI time-distance maps are plotted along the red-dashed lines of Figure \ref{fig1}f and Figure \ref{fig1}g, respectively. Panels (c) and (d) show the AIA 171 \AA\ intensity and HMI time-distance maps, respectively, of the jet of Figure \ref{fig2}. AIA 171 \AA\ intensity map (c) is plotted along the red-dashed line of Figure \ref{fig2}e, where HMI map (d) is plotted along the red-dashed line of Figure \ref{fig2}i. The green and white arrows, respectively, point to the jet-base brightening and minifilament. The yellow arrows in (b) and (d) point to the track of negative flux that converge towards the neutral line. 
	}  \label{fig2a}
\end{figure*}

 The EUI jet starts with a base-brightening that appears at the neutral line at 21:44:12 (Movie1a, green arrow in Figure \ref{fig1}b) and then a bright spire develops slowly (cyan arrow). The total duration of the jet is 7 minutes. Jet-base brightening (JBP) and jet spire can also be seen in the AIA 171 and 193 \AA\ images. AIA images also show a  cool-plasma structure at the jet-base region (white arrow in Figures \ref{fig1}e,i), which could be a signature of minifilament structure as it is rooted above the magnetic neutral line.  To see the evolution of the minifilament, we created  a time-distance map shown in Figure \ref{fig2a}a. We noticed that it does not erupt and it appears right next to the JBP.  
 However it is not clear whether the cool-plasma structure is a minifilament because (i) it is not an erupting structure as happens in typical coronal jets and (ii) the JBP usually appears underneath the erupting minifilament \citep{sterling15,panesar16b}. 
 

Co-aligned HMI magnetograms show that the small-scale jet occurs at the edge of the positive-polarity flux patch where negative polarity flux patch is also present next to it (see yellow arrows in Figure \ref{fig1}g). The negative polarity flux patch seems to converge towards the neutral line and cancels with the positive flux patch.  Figure \ref{fig2a}b shows the time-distance map of the jet-base region -- it shows a clear convergence of the negative-polarity flux patch towards the positive flux patch, which triggers the jet eruption. Although we are looking at the detection limit of the HMI magnetograms, it could be a possible signature of flux convergence and cancelation.

%

 \subsubsection{Example-2}

In Figure \ref{fig2}, we show another example of a small-scale coronal jet from our list (event-4 of Table \ref{tab:list}). The EUI images and the corresponding animation (Movie2a) show the jet that starts with a base brightening (green arrow) and followed by a jet spire (cyan arrow). This jet is also discernible in the AIA images (Figure \ref{fig2} and Movie2b). Interestingly, AIA images also show a nice minifilament structure (at 22:01:21 in Movie2b, white arrow in Figures \ref{fig2}a,d) at the base of the jet before and during the jet eruption. The minifilament starts to rise at 22:01:57 and then a base brightening appears where the minifilament was rooted prior to its eruption. The evolution of the minifilament and JBP can be seen in the time-distance map of Figure \ref{fig2a}c. It shows the clear rise of minifilament structure and the appearance of the base brightening underneath it. This scenario is similar to typical coronal jets, in which a coronal jet is driven by the eruption of a minifilament. This is an example of a small-scale coronal jet (base width 2000 $\pm$  225 km) that is driven by a minifilament eruption. The observed spire length (5000 $\pm$ 160 km) of this jet is shorter than our IRIS jetlet spire lengths \citep{panesar18b} and similar to the spire length of some of the Hi-C jetlets \citep{panesar19}. Whereas the observed jet-base width is shorter than the jet-base width of typical coronal jets \citep{panesar16b,panesar18a}. The jet propagates outward with an average speed of 58 $\pm$  0.5 \kms.

The pre-jet minifilament lies above the magnetic neutral line between the negative flux patch and positive magnetic flux patch (yellow arrows in Figure \ref{fig2}g). Both negative and positive flux patches are seen to converge towards the neutral line (Figure \ref{fig2}i).   The HMI time-distance map presents the scenario of flux convergence and cancelation at the jet-base region (Figure \ref{fig2a}d). Possible flux convergence and cancelation triggers the pre-jet minifilament eruption.

\vspace{1cm}
\subsubsection{Physical Properties}

We estimated the physical properties, e.g. jet duration, spire length, and spire width, of our small-scale jets using  EUI 174 \AA\ images (Table \ref{tab:list}). The duration of each jet was calculated when the jet-base brightening turns on until the spire fades away.  The spire length is measured from the base to the visible tip of the spire at the time of its maximum extent. The jet-base width is estimated at the widest part of the base during its peak brightening. 
The average duration of our five jets is 6.5 $\pm$ 4 minutes, which is typical for small-scale jets as also reported by \cite{hou21} using Solar Orbiter/EUI data. The average jet spire length comes to be 6050 $\pm$ 2900 km, which is smaller than the Hi-C 2.1 jet-like events \citep{panesar19}. Whereas the average jet-base width is 2200 $\pm$ 850 km that is similar to the base-width of EUI campfires \citep{panesar2021}. 
The average plane-of-sky speed of our jets is 60 $\pm$ 8  \kms, which is similar to the jet speeds reported by \cite{panesar18b,panesar19} and \cite{hou21}. The observational speeds are estimated using AIA 171 \AA\ images because, as previously mentioned, some of the EUI 174 \AA\ frames are binned and have variable compression settings. Also because the EUI 174 \AA\ images are not isochronal \citep{tiwari22}.

%
\begin{figure*}
	\centering
	\includegraphics[width=\linewidth]{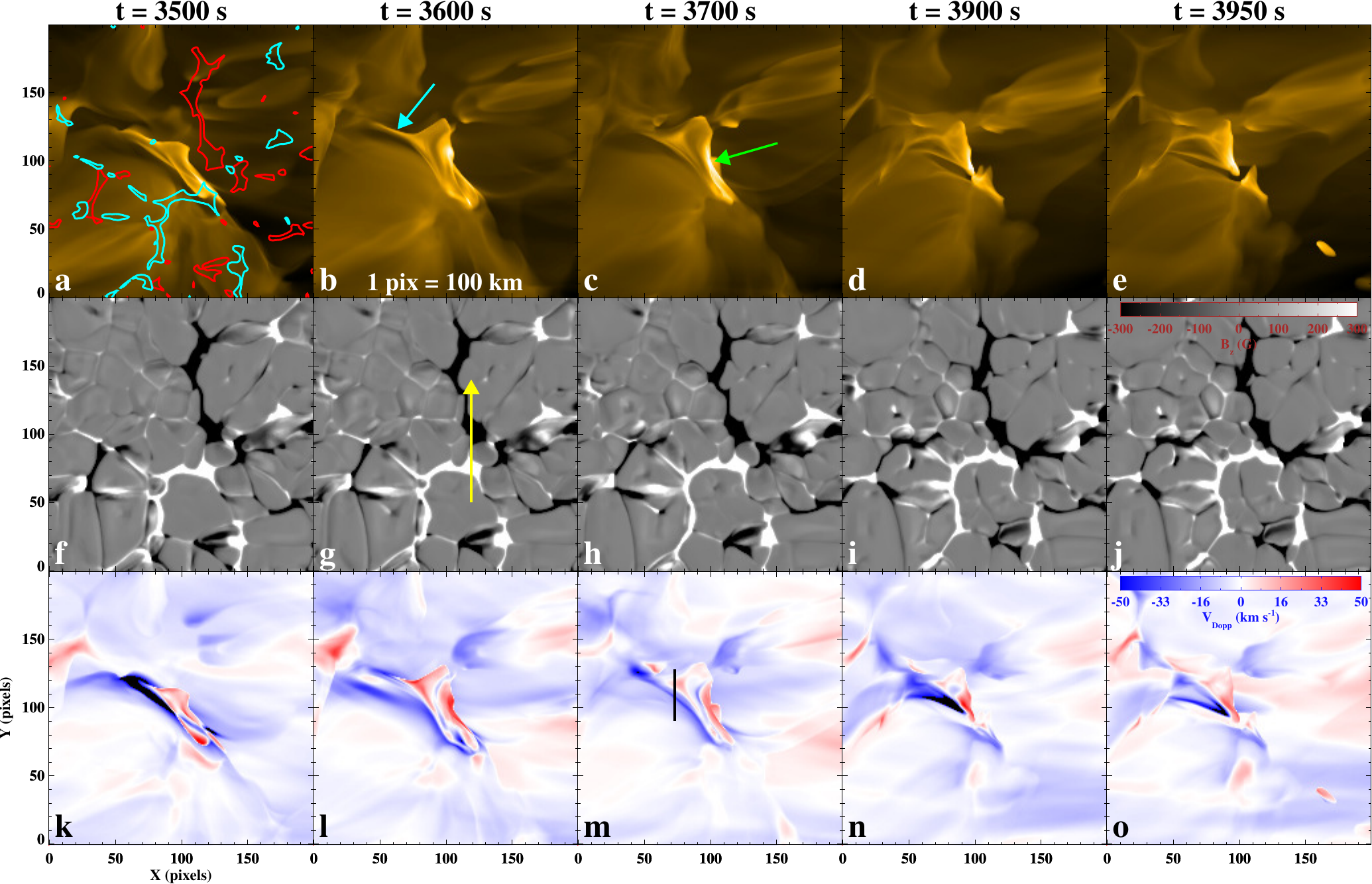}
	\caption{Coronal jet example-1 from the Bifrost MHD simulation (event-4 of Table \ref{tab:list}). Panels (a-e) show the jet in synthesized \FeIX/\FeX\ emission.  Panels (f-j) show the Bz map of the jet-base region. 
		Panels (k-o) display the \FeIX/\FeX\  Dopplershift, v$_{Dopp}$, with the upper and lower values saturated at 50 \kms. The contours, of levels $\pm$ 200 G, of Bz are overlaid on panel (a), where turquoise and red contours outline positive and negative magnetic flux, respectively.  The cyan and green arrows in (b and c) point to the jet spire and jet base brightening, respectively.  The yellow arrow in (g) shows the north-south cut along which the time-distance map is plotted and shown in Figure \ref{fig5}a. The black line in (m) shows the north-south cut for the  time-distance map of Figure \ref{fig6}a. The animation (Movie3) runs from 3250 s to 4300 s and the animation is unannotated. 
	}
	\label{fig3}
\end{figure*} 
%

\begin{figure*}
	\centering
	\includegraphics[width=\linewidth]{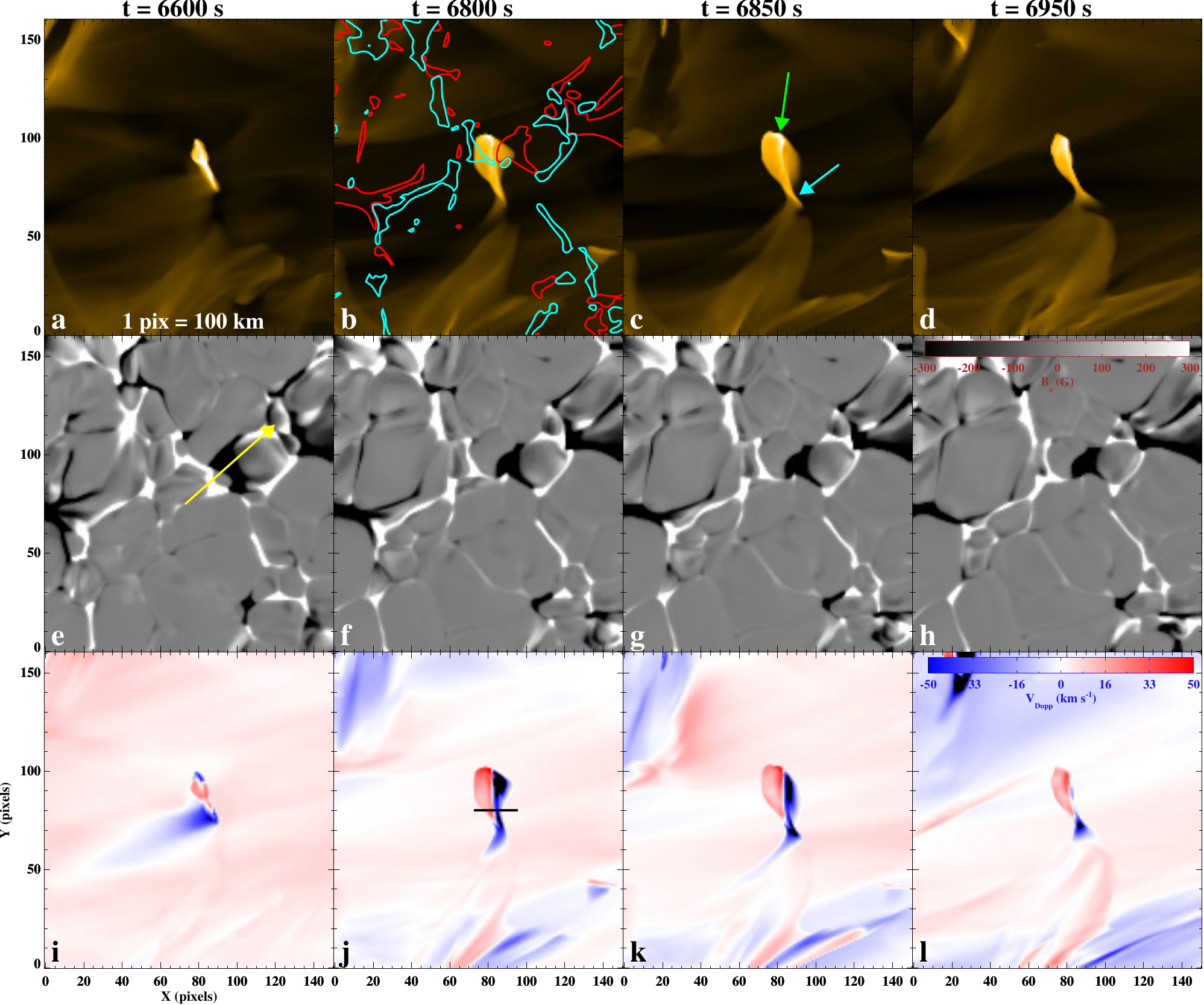}
	\caption{Coronal jet example-2 from the Bifrost MHD simulation (event-3 of Table \ref{tab:list}). Panels (a-d) show the jet in synthesized \FeIX/\FeX\ emission.  Panels (e-h) show the Bz map of the jet-base region. 
		Panels (i-l) display the \FeIX/\FeX\  Dopplershift, v$_{Dopp}$, with the upper and lower values saturated at 50 \kms. The contours, of levels $\pm$ 200 G, of Bz are overlaid on panel (b), where turquoise and red contours outline positive and negative magnetic flux, respectively.  The cyan and green arrows in (c) point to the jet spire and jet base brightening, respectively. The yellow arrow in (e) shows the diagonal cut along which the time-distance map is plotted and shown in Figure \ref{fig5}b. The black line in (j) shows the east-west cut for the time-distance map of Figure \ref{fig6}b. The animation (Movie4) runs from 6400 s to 7050 s and the animation is unannotated.
	} \label{fig4}
\end{figure*} 
%
%
%
%
\begin{figure}
	\centering
	\includegraphics[width=\linewidth]{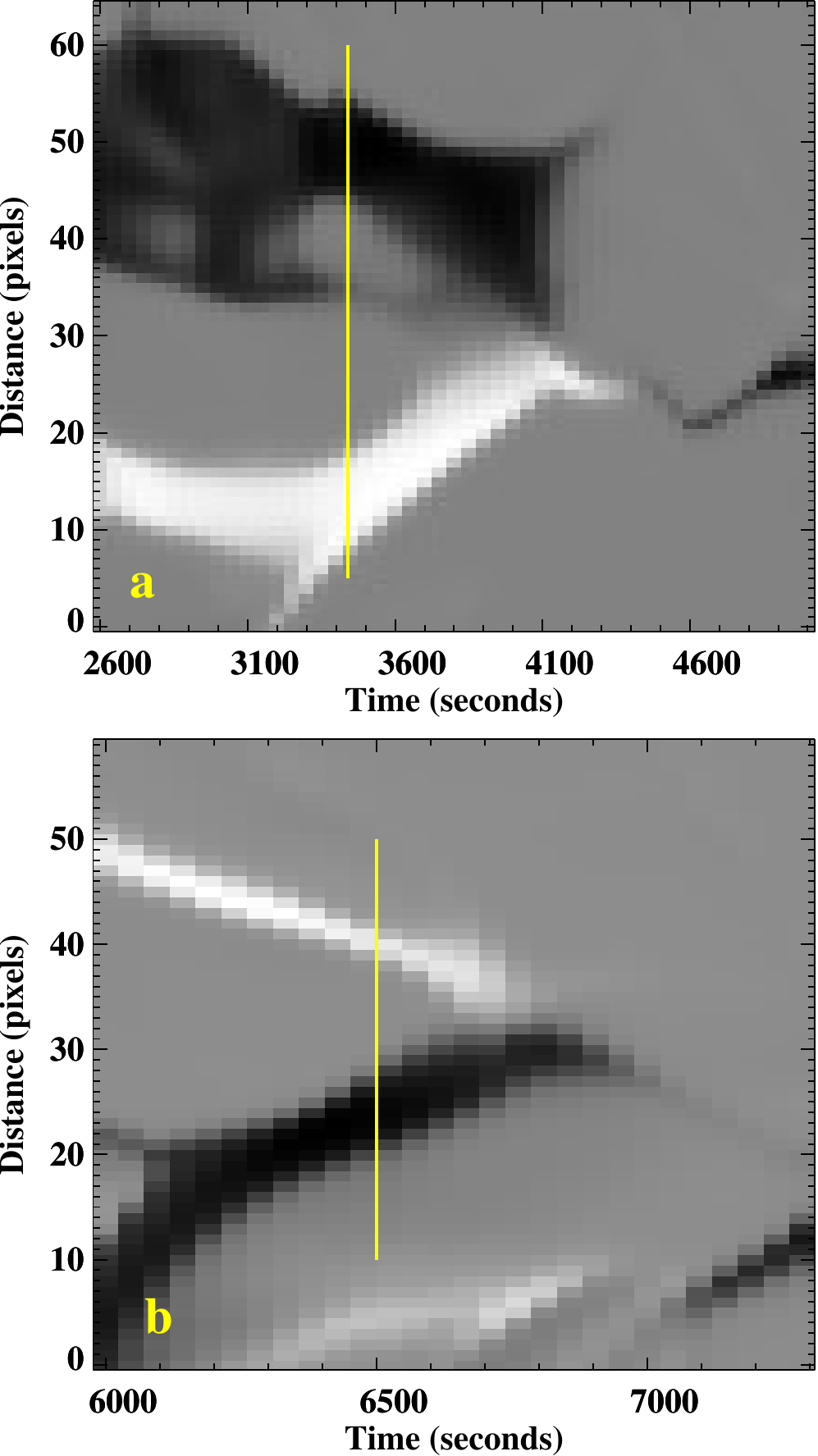}
	\caption{Magnetic field evolution. Panels (a) and (b)  show the Bz time-distance maps along the yellow arrows in Figures \ref{fig3}g and \ref{fig4}e, respectively. The solid yellow lines mark the onset of the jet that are shown in Figures \ref{fig3} and \ref{fig4}. The maps show the flux convergence and cancelation at the jet base region.
	} \label{fig5}
\end{figure} 
\subsection{Coronal jets in Bifrost MHD simulation} \label{bjet}

Using a Bifrost simulation, we synthesized  \FeIX\ and \FeX\ emissions to calculate line intensities, and Doppler speeds. We also analyze the vertical component of the magnetic field, Bz. This is the same set of simulation that has been analyzed by \cite{tiwari22} for exploring thermal, velocity, and magnetic properties of fine-scale bright dots. The simulation data is available for several hours but we analyzed the data for only two hours (starting simulation frame no. 420 to 576) at a 50 s cadence. The selected jets in our simulations have similar sizes as to the observed small-scale jets, selected intentionally for a fair comparison. 

 
 \subsubsection{Example-1}

In Figure \ref{fig3} and in corresponding animation (Movie3), we display our first example of a coronal jet from simulation (event-4 of Table \ref{tab:list}).  The upper panels (a--e) show the jet in synthesized \FeIX/\FeX\ emission. The jet shows a bright base (green arrow in  Figure \ref{fig3}c) and spire (cyan arrow in  Figure \ref{fig3}b) similar to the EUI jets. The duration of this jet is 14 min $\pm$  100 s, the jet-base width is 6600 $\pm$ 400 km, and the spire length is 7500 $\pm$  1900 km. These estimations are done in the same way as done for observations.

The jet resides above the magnetic neutral line between the positive- and negative-polarity flux patches (Figure \ref{fig3}a). The bright jet-base occurs at the neutral line similar to that of observations. 
We follow the evolution of the Bz maps with time and notice that positive and negative flux converge towards the neutral line. The flux cancelation between these two flux patches likely drives the jet. Further, to follow the evolution of jet-base magnetic field region more clearly, we created a Bz time-distance map (Figure \ref{fig5}a) along the yellow arrow of Figure \ref{fig3}g. The Bz time-distance map shows that positive and negative flux patches start to converge towards the neutral line before the jet onset. It continuously converges and cancels even after the jet eruption. The simulation shows a clear example of flux cancelation at the jet-base region and that flux cancelation leads to the jet. This behavior is in agreement with the observations of many coronal jets that are also seen to trigger with flux cancelation \citep[e.g.][]{shen12,adams14, panesar16b,panesar18a,sterling17,mcglasson19}. Furthermore, we noticed another jet (jet-5 of Table \ref{tab:list}) eruption from the same neutral line, which occurs due to the continuous flux cancelation. It is common to see sequential coronal jet eruptions from the same neutral line due to continuous flux cancelation \citep{panesar17,panesar18b} -- the jetting stops when one of the polarities fully disappears/cancels from the neutral line. Thus, at the end there is no neutral line and there are no more jets. 

Figures \ref{fig3}(k--o) display the \FeIX/\FeX\ Doppler velocity maps of the jet. The Dopplergrams show downflows (redshifts) at the base of the jet. Whereas the two opposite Dopplershifts (blueshifts and redshifts) were seen on the two  edges of the jet spire (Figure \ref{fig3}m). The opposite Dopplershifts in the jet spire are often a signature of the untwisting motion of the magnetic field in the jet spire \citep[e.g.][]{pike98,kamio2010,curdt2012,cheung15,tiwari18,sterling19,panesar22}. The redshifts at the base of the jet is considered to be a signature of downflows at the location of  jet bright point (in the reconnected closed-loops) \citep{tiwari19,panesar22}. 

To present a quantitative picture of Doppler flows along the jet spire, we make the Doppler speed plots. For that purpose, we take a cut across the jet spire (black line in Figure \ref{fig3}m) and plot their Doppler speeds in Figure \ref{fig6}a. The jet spire shows stronger blueshifts reaching up to 35 \kms\ and redshifts peaking at 10 \kms. Similar but stronger (up to 75 \kms) Doppler flows have  been reported by \cite{tiwari22} in bright dots of \FeIX/\FeX\ emission.

 \subsubsection{Example-2}

In Figure \ref{fig4}, we present second example of a coronal jet from our simulation (event-3 of Table \ref{tab:list}). The corresponding animation (Movie4) shows the same jet over the duration of the event.  This is the smallest jet from our simulation. The jet starts with a base brightening (green arrow) and a spire (cyan arrow) that expands with time. Jet becomes maximum in size at about 6850s (Figure \ref{fig4}c). After that it begins to fade away. The lifetime of the jet is 9 min $\pm$ 50 s. It has a base width of 1770 $\pm$ 80 km, and spire length of 4130 $\pm$ 120 km. 

In agreement with the above example, this jet occurs at the neutral line and the jet bright point straddles above the neutral line (Figure \ref{fig4}b). The Bz maps show a clear scenario of flux convergence and cancelation at the jet-base region (Figure \ref{fig4}e--h), which leads to the jet. In Figure \ref{fig5}b, we show a Bz time-distance map that is created along the diagonal arrow of Figure \ref{fig4}e to display the magnetic field evolution of the jet-base region. One can see that both the flux patches clearly converge and cancel and that flux cancelation triggers the jet. The positive flux patch has completely been eaten-up by the negative flux patch. Hence no more jets occur from this location. 

\begin{figure}
	\centering
	\includegraphics[width=\linewidth]{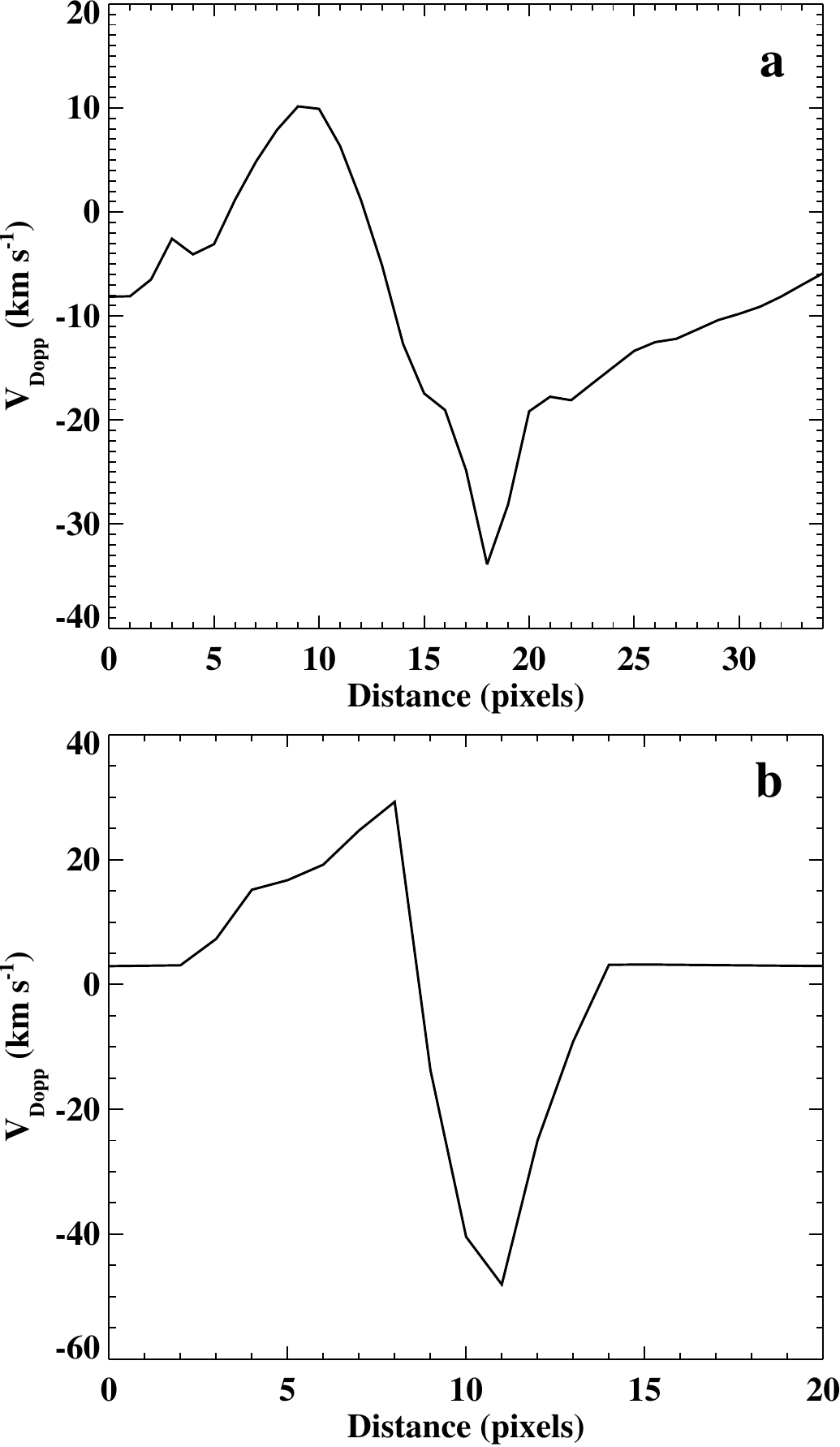}
	\caption{Doppler velocities of the jet spire. Panels (a) and (b) show the Doppler profiles of the jet spire along the black lines in Figures \ref{fig3}m and \ref{fig4}j, respectively. The profiles clearly display the upflows (blueshifts) and downflows (redshifts) in the jet spire. In these plots, 1 pixel corresponds to 100 km.
	} \label{fig6}
\end{figure}

The \FeIX/\FeX\ Doppler velocity maps of this jet example is shown in Figure \ref{fig4}(i--l). In the beginning phase of the jet, dominant redshifts are observed. Later, both redshifts and blueshifts are seen in the jet structure (Figures \ref{fig4}(j,k)). The blueshift and redshift next to each other is consistent with the untwisting motions as seen in typical jet observations. For example, in Figures \ref{fig4}(j,k), the extended blueshift in the south-east direction corresponds to the jet spire and the redshift and blueshift at the base extending to the jet spire corresponds to untwisting motion.  Doppler speed plot is shown in Figure \ref{fig6}b that is made along the horizontal line of Figure \ref{fig4}(j). It shows the quantitative behavior of the Doppler flows in jet spire. The plot shows the stronger blueshifts in the jet spire that reaches up to 50 \kms, whereas redshift goes up to 30 \kms. Similar but weaker Doppler flows have been reported by \cite{panesar22} in a quiet region coronal jet captured by \iris\ in \MgII\ k spectra.

\subsubsection{Temperature maps}

As mentioned in Section \ref{data}, in our simulation there is a possibility of absorption from neutral gas due to the shorter wavelength of iron lines. Therefore, we also compute the intensities of  \FeIX\ and \FeX\ lines with absorption turned on. This will give us an idea if there is any chromospheric cool gas/plasma present at heights of $>$ 2 Mm.

\begin{figure*}
	\centering
	\includegraphics[width=\linewidth]{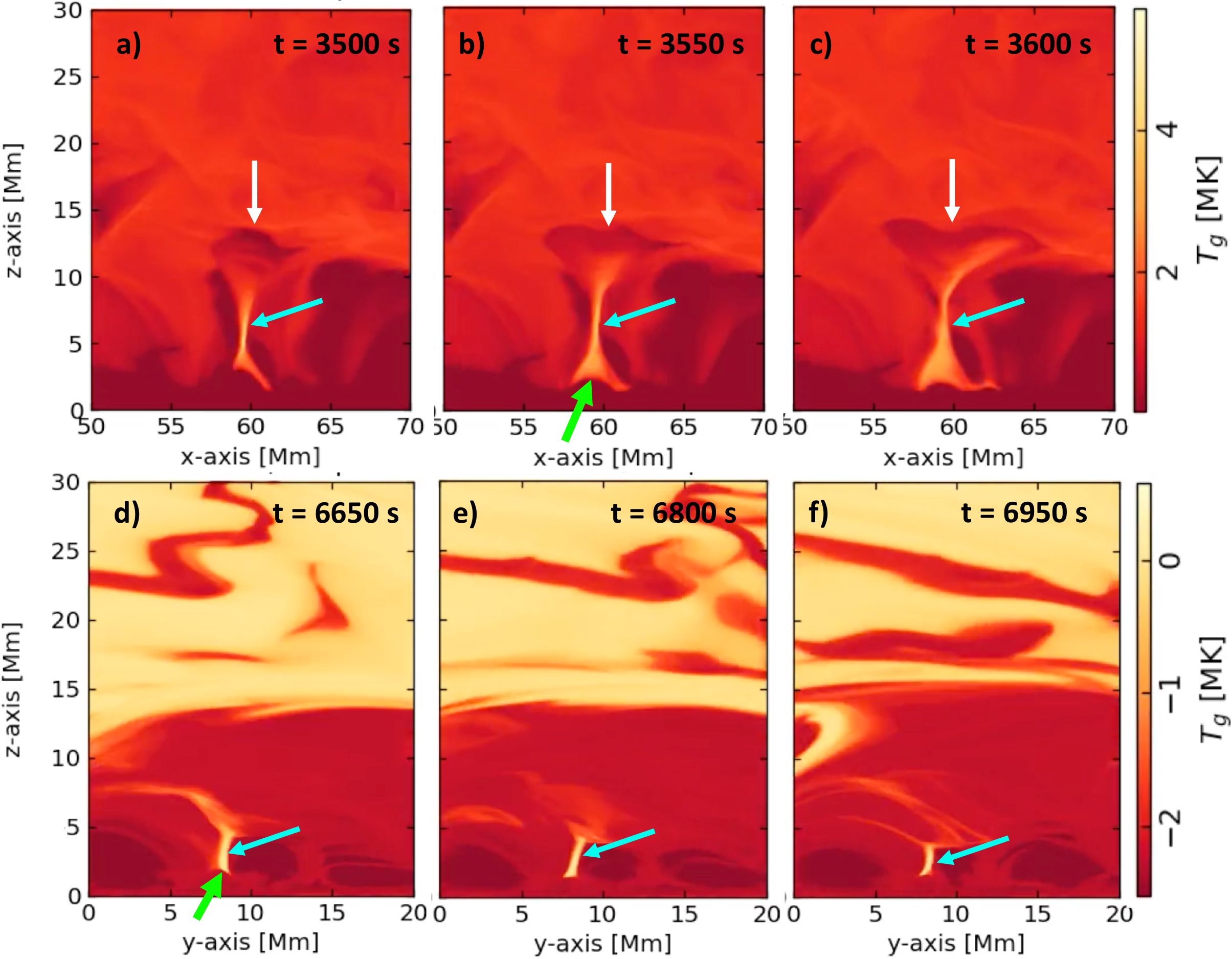}
	\caption{Temperature maps. Panels (a--c) and (d--f), respectively, show the temperature maps for coronal jet of Figures \ref{fig3} and \ref{fig4}. These maps are created across the jet structure versus height (in z-direction). The z-axis corresponds to height above the photosphere (z=0).  The temperature is logarithmic in units of MK, where 1 MK is zero, 0.1 MK is 100 000 K or -1, and so on. The ‘g' stands for gas in Tg. The colorbar scale in the top row, panels (a-c), is linear, while the colorbar scale in the bottom row, panels (d-f) is logarithmic. The white arrows points to the cool-plasma, cyan arrows point to the jet spire, and green arrows point  to the jet-base region. 
	} \label{fig7}
\end{figure*}

Figures \ref{fig7}(a--c) display the temperature maps of the jet of Figure \ref{fig3}, at different times. The jet spire and jet-base are shown by cyan and green arrows, respectively. These maps also show the presence of chromospheric cool-plasma during the eruption onset. The cool-plasma (white arrows) can be seen on the top of the jet spire when the jet is moving outwards. This scenario is in agreement of our EUI and AIA jet observation, where we also observe erupting cool-plasma (minifilament) during the jet onset (Figure \ref{fig2}). This cool-plasma structure has some similarities with the cool-plasma previously seen by \cite{hansteen19} in simulations of UV bursts. However, the reconnection that happens in our model, on large scales, is between newly emerged field and ambient coronal field, so there is not as much “cool” material in between the observer and the reconnection site as there was in \cite{hansteen19}. There was 5 to 10 Mm of cold ($T < 10^4$ K) high opacity material between the reconnection location (at low to moderate heights, 0 -- 2 Mm above the photosphere) between two different bubbles of the emerging field and the corona in \cite{hansteen19}.

In Figures \ref{fig7}(d--f), we show the temperature maps for the jet of Figure \ref{fig4}. This is the smallest jet example of our simulation and we can see that it does not go high in the z-axis. The jet spire and jet-base are pointed with an arrow. Unlike the previously described jet, this jet does not show the presence of cool-plasma. One possibility is that the jet size is so small therefore, it is hard to detect the cool-plasma in it. Another reason is that may be there is no cool-plasma present in this jet. 

We investigated five jets in simulation and find that most of them are similar to the observations. Similar in the sense that four out of five jets occur at the neutral line where flux cancelation takes place and four out of five jets confirm the presence of erupting cool-plasma. Their durations, spire-lengths, base-widths, and speeds are also more or less in the range of observed EUI jets (see Table \ref{tab:list}).

%

\begin{figure}
	\centering
	\includegraphics[width=0.48\textwidth \vspace{0.1cm}]{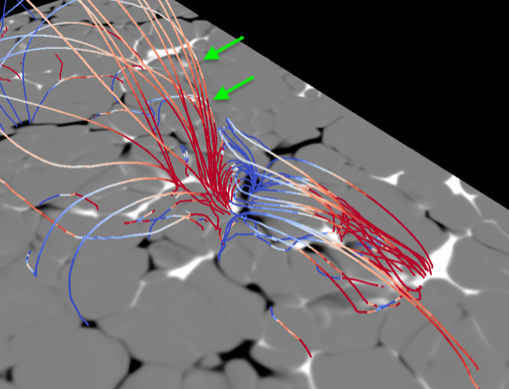} 
	
	\includegraphics[width=0.48\textwidth]{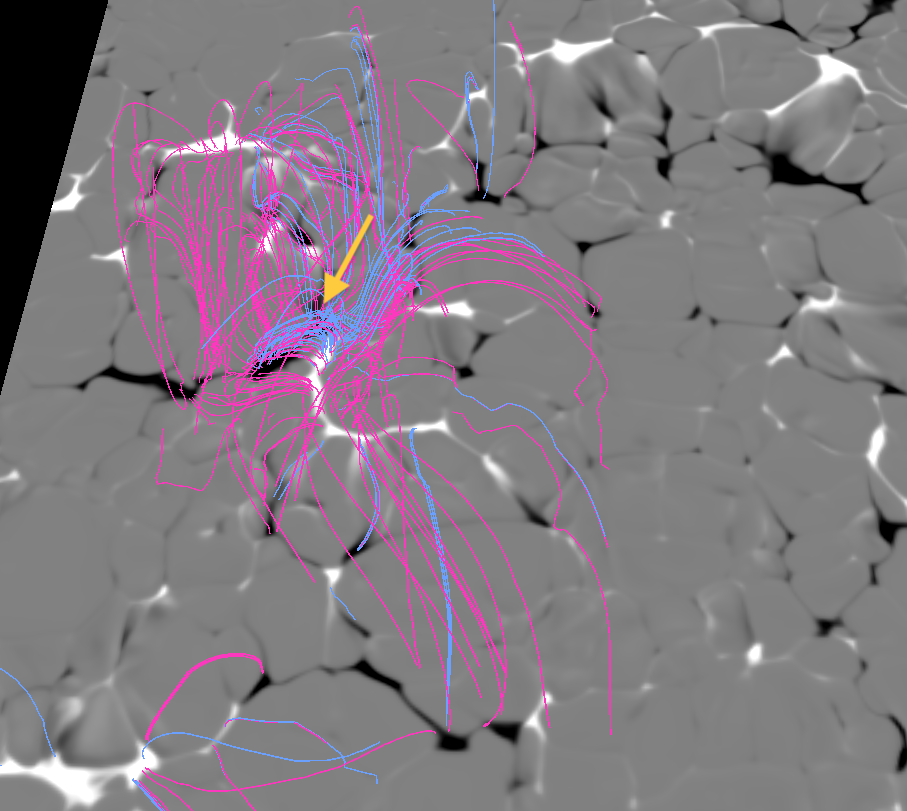} 
	\caption{Magnetic field topology of the jet shown in Figure \ref{fig3}. The upper panel shows 3D magnetic field geometry for the jet. The background image is Bz map that is saturated at $\pm$500 G. The red and blue lines in upper panel represent the positive and negative magnetic fields, respectively. The green arrows point to the newly-reconnected far-reaching magnetic field lines along which the jet material escapes. The bottom panel shows magnetic field topology from a different angle during the jet onset (t=3500 s). The orange arrow points to the low-lying closed loops, near the location of the jet bright point. We used a different color combination in the bottom panel for drawing field lines for a better visibility of the loops pointed to by the orange arrow.
	}  \label{fig8}
\end{figure} 

\subsubsection{Magnetic field topology}

We used the visualization software VAPOR \citep{li-vapor19}, to trace the magnetic field lines of our jets. 3D magnetic field geometry will help us to understand  the jet formation and eruption mechanism. 

Here we  show the magnetic field  topology of jet-4 of Table \ref{tab:list} (Figure \ref{fig8}). In the upper panel of Figure \ref{fig8}, the red and blue extrapolated loops, respectively, represent the positive and negative photospheric magnetic field. The green arrows point to the far-reaching magnetic field along which the jet material escapes outward and appears as a bright jet spire, in the same direction of far-reaching magnetic field, in the synthetic \FeIX/\FeX\ images of Figure \ref{fig3}b.  The bottom panel shows the magnetic field geometry during the jet onset, at t=3500 s, at a relatively different angle than the above image. This facilitates us to see the different set of magnetic field lines at different angles. The orange arrow points to the closed field lines that are anchored above the neutral line. These closed field lines are the lower product of magnetic reconnection and a site of jet base brightening \FeIX/\FeX\ images of Figure \ref{fig3}c. Jet bright point often appears at the location of an erupting minifilament (green arrows in Figures \ref{fig2}e and \ref{fig3}c). It is a miniature version of solar flare arcade where downflows are often dominant \citep{imada13}.

%
%
\vspace{1cm}
\section{Discussion}
We selected five quiet Sun region small-scale coronal jets using \hri\ images, and investigated their origin using SDO/AIA and SDO/HMI data. In addition, we compared the observed jets with five analogous small-scale jets that we found in a Bifrost MHD simulation of a quiet Sun region. Overall, all five jets from the simulation show similarities with the jets from \hri\ and SDO/AIA observations. All of these jets show resemblance with our earlier findings of typical coronal jets. The main difference is the size scale. The average base-width of EUI jets is 2200 $\pm$ 850 km, which is much smaller than the average base-widths of typical coronal jets (12000 km; \citealt{panesar18a}). The base widths of our EUI jets are similar to the sizes of the jets reported by \cite{hou21}. Some of our jets may be considered as`jet-like' campfires reported by \cite{panesar2021}, but these are different from dot-like and loop-like campfires \citep{berghmans2021,tiwari22} because these jets are much bigger elongated structures moving outwards with enhanced base brightenings. 

The sizes of our EUI jets are not very different from the sizes of our jets in the Bifrost simulation. On average, the spire length of the EUI jets is 6050 $\pm$ 2900 km whereas the jets in the simulation have spire length of 6500 $\pm$ 2900 km, which are quite similar. The observed spire lengths are smaller than the lengths of IRIS jetlets (16000 km; \citealt{panesar18b}), Hi-C jets (9000 km; \citealt{panesar19}), and EUI micro jets (7700 km; \citealt{hou21}). The base-width of our EUI jets (2200 $\pm$ 850 km) is a little bit smaller than the base-width of simulated jets (3900 $\pm$ 2100 km) and IRIS jetlets (4400 km; \citealt{panesar18b}). But it is larger than the base-widths of campfires (1600 km; \citealt{panesar2021}). The average speed of our jets in observations (60 $\pm$ 8 \kms) is in the same order of that for the jets in simulation (42 $\pm$ 20 \kms). On average, the simulated jets are a little bit slower than in observations. This could be due to the limited height of the simulated box.
The average lifetime of our small-scale jets is 6.5 minutes, which is somewhat similar to the duration of simulated jets (9.0 minutes) and two times longer than that of  IRIS jetlets (3 minutes; \citealt{panesar18b}).  The temporal cadence of the simulation is 50 s as compared to that of 10 s in the EUI observations, resulting into the selection of the jets with relatively longer lifetimes in our simulation. That means, statistically, the observed and simulated jets perhaps have very similar lifetimes. 

The reported jets here show qualitative similarities with typical coronal jets, and thus are ``small-scale" coronal jets. All our jet physical parameters (e.g. spire lengths, base widths, and lifetimes) are significantly smaller than those for typical coronal jets, on average. As discussed above the observed parameters of jets in simulation are almost similar to the parameters of EUI jets. This gives us an opportunity to better understand the dynamics small-scale coronal jets in high resolution simulations.

Essentially all of our EUI jets originate from magnetic neutral lines. Three out of five jets occur at clear site of flux cancelation, consistent with the findings of \cite{panesar16b} for typical quiet Sun jets. In two events, we noticed that opposite flux is present at the jet-base region but its convergence towards neutral line and then cancelation is not clear enough. This might be due to the fact that we are looking at the detection limit of HMI magnetograms. 

The Bifrost simulation plays an important role in our study as it clearly shows the photospheric evolution of these small-scale jets, and the plasma flows in them. Four out of five jets occur at neutral lines where both the opposite polarity flux patches were present and the jets occurred during the process when positive and negative flux patches converged towards the neutral line, reconnect, and cancel with each other. We surmise that this is a reconnection driven flux cancelation where flux cancelation is a consequence of the submergence of the lower product of the reconnected loops into the photosphere \citep{balle89,priest21,syntelis21,hassanin22}; a clear example of this mechanism can be found in \cite{tiwari14}. 
This is apparently in agreement with our present and earlier jet observations that magnetic reconnection driven flux cancelation, between majority-polarity flux and a merging minority-polarity flux patch, is the main cause of the formation and triggering mechanism of many coronal jets \citep{panesar16b,panesar18b}. To the best of our knowledge, this is the first time that coronal jets are reported to systematically accompany magnetic flux cancelation in any  3D MHD simulation.

However, one jet in our simulation comes from the location where no magnetic field was present. Probably, there is some other mechanism that leads to this jet. 
In the simulation, we also notice that flux convergence and cancelation often occur at the edges of network regions and jets appear at some of these locations. We interpret that to have a coronal jet first we need to have a shear field and flux rope present at the neutral line. \cite{panesar17} found that pre-jet minifilaments mainly appear above sheared canceling neutral lines. Thus, no jets with minifilaments can occur at a neutral line in a  potential bipolar magnetic field configuration. The above mentioned scenario is similar to the formation and eruption mechanism of typical solar filaments \citep{balle89,moore92}. Nonetheless flux cancelation plays a significant role in the formation of small-scale jets as  evident from our analyzed jets from the observations and simulation. 

The magnetic field geometry of our jets in the Bifrost MHD simulation shows the formation of a small-scale loop at the neutral line during the jet. We infer that this small-scale loop is the lower product of magnetic reconnection as it appears above neutral line during the jet onset. The jet base brightening often appears at the neutral line during the jet onset \citep{sterling15}. 

Apart from one event, the remaining four jets in the Bifrost simulation show the evidence of chromospheric cool-gas/plasma. The cool plasma is seen when the jet is lifting off and the jet spire shows the signature of both cool and hot plasma during the eruption, as previously noted in observations by \cite{moore13}. The erupting cool plasma is the evidence of twisted flux rope that might be present at canceling neutral line.  This is in agreement with our \hri\ and AIA jet observations where in three out of five cases a minifilament structure is seen.  Some simulations have discussed the formation of flux rope in typical coronal jets \citep[e.g.][]{wyper18,doyle19}.

The opposite Doppler shifts (redshift and blueshift) have been seen in all the five jets in the simulation, which is an additional evidence of the untwisting motion in the magnetic field of the jet spire. The opposite Doppler shifts have been captured by the \iris\ spectra in active region jets \citep[e.g.][]{cheung15,lei19,ruan19,tiwari19,zhang21}. Recently, opposite Doppler shifts  in an on-disk quiet region coronal jet have been observed by \iris\ in \MgII\ k spectra \citep{panesar22}. The untwisting motions in the spire of coronal jets have also been reported in EUV observations \citep[e.g.][]{moore10,schmieder13,panesar16a}. 

We note that the structure of jets evolve and change on the order of tens of seconds, which for their length scales means single slit rasters would be too slow to follow their evolution. Therefore, the characteristics of small-scale jets can be studied using an integral field spectrometer such as MUSE \citep{pontieu22,cheung22}. The coordinated observations from DKIST, the Goode Solar Telescope, and SST together with MUSE would provide more information on the formation of small-scale coronal jets.

Our observations and simulation of small-scale jets show that jets frequently occur at the edges of quiet Sun network regions and their formation and eruption mechanisms are similar to those of typical coronal jets. 
 \hri\ data gives us an opportunity to study and better understand the formation and eruption mechanism of small-scale jets. In future, we will explore the formation mechanism of those jets that are not driven by minifilament eruption and are not visibly triggered by flux cancelation. Bifrost simulation of jets, with a larger spatial simulation domain and better cadence, will shed more light on the formation mechanism of coronal jets.

\section{Conclusions}
We analyze small-scale coronal jets using EUI 174 \AA\ observations, and SDO data, and compare their properties with similar jets found in a Bifrost MHD simulation. Both the observations and simulation show a consistent picture of coronal jets in that majority of them form by magnetic reconnection accompanying flux cancelation at the neutral line. They also show the evidence of cool-plasma structure. These findings are in agreement with the observations of typical coronal jets and larger-scale CME-producing eruptions. The observed small-scale jets most likely are the miniature versions of them.

\vspace{0.3cm}

NKP acknowledges support from NASA's SDO/AIA (NNG04EA00C) and HGI grant  (80NSSC20K0720).  SKT gratefully acknowledges support by NASA HGI grant (80NSSC21K0520) and NASA contract NNM07AA01C (Hinode).
We acknowledge the use of  Solar Orbiter/EUI and  \sdo/AIA/HMI data. AIA is an instrument onboard the Solar Dynamics Observatory, a mission for NASA’s Living With a Star program. Solar Orbiter is a space mission of international
collaboration between ESA and NASA, operated by ESA. The
EUI instrument was built by CSL, IAS, MPS, MSSL/UCL,
PMOD/WRC, ROB, LCF/IO with funding from the Belgian
Federal Science Policy Office (BELSPO/PRODEX); the
Centre National d’Etudes Spatiales (CNES); the UK Space
Agency (UKSA); the Bundesministerium für Wirtschaft und
Energie (BMWi) through the Deutsches Zentrum für Luft- und
Raumfahrt (DLR); and the Swiss Space Office (SSO). This work has made use of NASA ADSABS and Solar Software.


\bibliographystyle{aasjournal}

\end{document}